\documentclass[12pt]{article}
\usepackage{amsfonts}
\usepackage{graphicx}
\usepackage{amssymb}
\usepackage{amsmath}
\def \be {\begin{equation}}
\def \ee {\end{equation}}
\def \ba {\begin{array}}
\def \ea {\end{array}}
\def \bea{\begin{eqnarray}}
\def \eea{\end{eqnarray}}

\def \be {\begin{equation}}
\def \ee {\end{equation}}
\def \ba {\begin{array}}
\def \ea {\end{array}}
\def \bea{\begin{eqnarray}}
\def \eea{\end{eqnarray}}

\def \a {\alpha}

\def \m {\mu}
\def \n {\nu}
\def \k {\kappa}

\def \L {\Lambda}
\def \s {\sigma}

\def \r {\rho}

\def \th {\theta}

\def \f {\frac}

\def \nn {\nonumber}

\def \sr {\sqrt}

\def \wt {\widetilde}

\def \lab {\label}

\begin{document}
\renewcommand{\thefootnote}{\fnsymbol{footnote}}
\begin{titlepage}

\vspace{10mm}
\begin{center}
{\Large\bf Thermodynamic approach to field equations in Lovelock gravity and $f(R)$ gravity revisited}
\vspace{16mm}

{\large Yan-Gang Miao${}^{1,2,}$\footnote{E-mail address: miaoyg@nankai.edu.cn},
Fang-Fang Yuan${}^{1,}$\footnote{ffyuan@nankai.edu.cn}, and
Zheng-Zheng Zhang${}^{1,}$\footnote{zhangzhengzheng@mail.nankai.edu.cn}

\vspace{6mm}
${}^{1}${\normalsize \em School of Physics, Nankai University, Tianjin 300071, China}

\vspace{3mm}
${}^{2}${\normalsize \em State Key Laboratory of Theoretical Physics, Institute of Theoretical Physics, \\Chinese Academy of Sciences, P.O. Box 2735, Beijing 100190, China}}


\end{center}

\vspace{10mm}
\centerline{{\bf{Abstract}}}
\vspace{6mm}
\noindent
The first law of thermodynamics at black hole horizons is known to be obtainable from the gravitational field equations.
A recent study claims that
the contributions at inner horizons should be considered in order to 
give the conventional first law of black hole thermodynamics.
Following this method,
we revisit the thermodynamic aspects of field equations in the  Lovelock gravity and $f(R)$ gravity
by focusing on two typical classes of charged black holes in the two theories.

\vskip 20pt
\noindent
{\bf PACS Number(s)}: 04.50.Kd, 04.70.Dy


\end{titlepage}
\newpage
\renewcommand{\thefootnote}{\arabic{footnote}}
\setcounter{footnote}{0}
\setcounter{page}{2}

\section{Introduction}

Given the action of a gravitational system, the field equations can be obtained through the variation with respect to a metric.
Although the field equations ought to contain all available dynamical information of the system including some thermodynamic properties, 
it is still surprising that through a quite simple procedure a relevant component of the field equations in a static and spherically symmetric spacetime can be rewritten~\cite{Padmanabhan:2002sha} as the form of the first law of thermodynamics at a black hole horizon.
Since then this approach has been applied to various black hole solutions
including the ones in the Lovelock gravity~\cite{Paranjape:2006ca} and $f(R)$ gravity~\cite{Akbar:2006mq}.

As noted in e.g. ref.~\cite{Kothawala:2009kc}, the resulting expression is actually different from the conventional first law of black
hole thermodynamics because of the appearance of the $P d V$ term, where $P$ is the radial pressure and $V$ is the volume surrounded by a horizon.  
It has been shown~\cite{Akbar:2007qg,Kwon:2013dua} for the BTZ black holes in several gravity theories that the expected first law (with the variation of black hole charges)
can be obtained by summing the contributions of the first law of thermodynamics at all black hole horizons.

Based on the idea of ref.~\cite{Kwon:2013dua}, we reconsider the thermodynamic aspects of field equations in the Lovelock gravity and $f(R)$ gravity, respectively.
In contrast with the general analysis in refs.~\cite{Paranjape:2006ca,Akbar:2006mq},
the procedure of ref.~\cite{Kwon:2013dua} involves the explicit 
relations between the positions of horizons and black hole charges.
Due to the fact that the black hole thermodynamics in the Lovelock gravity and $f(R)$ gravity has been studied intensively,
it is interesting and meaningful to compare the results obtained from the idea of ref.~\cite{Kwon:2013dua} with that from the general one.
Therefore, we choose two typical classes of charged black holes in the two gravity theories for our investigations.

This paper is organized as follows.
In the next section, we  demonstrate how to derive the first law of black hole thermodynamics for a particular class of charged Lovelock black holes in the 5-dimensional spacetime.
In section \ref{sec3}, we discuss a class of 4-dimensional $f(R)$-Maxwell black holes imposed by a constant curvature scalar, where this class of black holes has four horizons.
The conclusion 
is given in the last section.

\section{Lovelock black holes: 5-dimensional case}

For simplicity, we start with a class of charged black holes in Einstein-Maxwell-Gauss-Bonnet theory~\cite{Wiltshire:1985us,Charmousis:2008kc}.
Obviously more general static Lovelock black holes can be studied along the same way.

The action we consider here has the following form,
\be
S = \f{1}{16\pi G} \int d^d x \sr{-g} \big( R - 2\L + \a L_{GB} \big) + S_m,
\ee
where $L_{GB} \equiv R^2 - 4 R_{\m\n} R^{\m\n} + R_{\m\n\r\s} R^{\m\n\r\s}$ is the Gauss-Bonnet Lagrangian, the matter part $S_m$ is assumed to be the Maxwell term, and the parameter $\a$ is the Gauss-Bonnet coupling.

When the spacetime dimension $d=5$ and the cosmological constant $\L=0$, the corresponding black hole solutions with spherical horizons have been found~\cite{Wiltshire:1985us} to be
\be
ds^2 = - V(r) dt^2 + \f{1}{V(r)} dr^2 + r^2 \bigg[ \f{1}{1-\chi^2} d\chi^2 + \chi^2 ( d\th^2 + \sin\th^2 d\phi^2 ) \bigg],
\ee
\be \label{ncet}
V(r) = 1 + \f{r^2}{4\a} \Bigg[ 1 - \sr{1+ 8\a \Big(\f{2\m}{r^4} - \f{q^2}{r^6} \Big)} \Bigg], \qquad F_{tr} = \f{q}{4\pi r^3},
\ee
where $F_{tr}$ is the nonzero component of electromagnetic tensors and $\m$ is a parameter related to the black hole mass.
The horizon positions  are at
\be
r_\pm^2 = \m-\a \pm \sr{(\m-\a)^2 - q^2},
\ee
from which we can find the following relations:
\be   \lab{lre}
(r_+ r_-)^2 = q^2, \qquad r_+^2 + r_-^2 = 2 (\m - \a).
\ee

Based on the analysis of ref.~\cite{Paranjape:2006ca}, we arrive at the $(r,r)$ component of field equations at the outer horizon,
\be   \lab{lrr}
- \f{4\a q^2 + r_+^3 \Big(r_+^3 - \sr{r_+^6 - 8\a q^2 + 16\a\m r_+^2} \Big)}{2\a r_+^3 \sr{r_+^6 - 8\a q^2 + 16\a\m r_+^2}} (r_+^2 + 4\a)
-2
= \f{2}{3} r_+^2 \cdot 8\pi G P_+,
\ee
where $T^{\mu \nu} = F^{\mu\lambda} F^\nu_{\ \lambda} - \frac{1}{4} g^{\mu\nu} F^{\lambda\sigma} F_{\lambda\sigma}$ is the energy-momentum tensor and $P_+ \equiv T^r_{\ r} (r_+) $  is the radial pressure. An analogous equation at the inner horizon can be obtained easily by the replacement of $r_+$ by $r_-$ in eq.~(\ref{lrr}).

Since $V(r_+) = 0$, we have $\sr{r_+^6 - 8\a q^2 + 16\a\m r_+^2} = r_+ (r_+^2 + 4\a)$.
Considering the relation $(r_+ r_-)^2 = q^2$, we can rewrite the first term of eq.~(\ref{lrr}) to be $2 \Big(1-\f{r_-^2}{r_+^2} \Big)$.
Noting that the volume in this case is $V_+ = \f{S_3}{4}r_+^4 = \f{\pi^2}{2} r_+^4$,
we multiply by $\f{3\pi}{8 G} r_+ dr_+$ on both sides of eq.~(\ref{lrr}) and get a simpler form,
\be
T_+ d S_+ - d E_+ = P_+ d V_+,
\ee
with
\be
T_\pm = \f{1}{2\pi} \f{r_+^2 - r_-^2}{r_\pm(r_\pm^2 + 4\a)}, \qquad
S_\pm = \f{\pi^2}{2G} r_\pm^3 \Big(1+\f{12\a}{r_\pm^2}\Big), \qquad
E_\pm = \f{3\pi}{8G} (r_\pm^2 + 2\a).
\ee
This is the expected first law of thermodynamics, where $E_\pm$ coincides with the Misner-Sharp energy inside the horizons.

More explicitly, we have the corresponding equations at the outer and inner horizons as
\bea
\f{3\pi}{4G} \Big( \f{r_+^2 - r_-^2}{r_+} d r_+ - r_+ d r_+ \Big) &=& P_+ d V_+,  \lab{lex 1} \\
\f{3\pi}{4G} \Big( \f{r_-^2 - r_+^2}{r_-} d r_- - r_- d r_- \Big) &=& P_- d V_-.  \lab{lex 2}
\eea
Following the procedure described in ref.~\cite{Kwon:2013dua}, the sum of these two equations reads
\be
- \f{3\pi}{4G} (r_+ dr_+ + r_- dr_-) + T_+ d S_+ + \f{3\pi}{4G} \Big( r_- - \f{r_+^2}{r_-} \Big)d r_-
= P_+ d V_+ + P_- d V_-. \label{sum}
\ee

From the relations in eq.~(\ref{lre}), we have
\begin{eqnarray}
d \m - d\a &=& r_+ dr_+ + r_- dr_-, \nonumber \\
q d q &=& r_+ r_- ( r_+ d r_- + r_- d r_+ ). \label{dmaq}
\end{eqnarray}
Thus, if we recall the parameters, i.e., the black hole mass $M$, the total electric charge $Q$, and the conjugate potential $\Phi_+$,
\be
M = \f{3\pi}{4G} \m, \qquad Q = \Big(\f{\pi}{4G}\Big)^{\f{2}{3}} q, \qquad \Phi_+ = 3 \Big(\f{\pi}{4G}\Big)^{\f{1}{3}} \f{Q}{r_+^2},
\ee
eq.~(\ref{sum}) can be rewritten as
\be     \lab{m2}
- d M + \f{3\pi}{4G} d \a + T_+ d S_+ + \Phi_+ d q  
- \f{3\pi}{4G} \Big( \f{r_-^2}{r_+} d r_+ + \f{r_+^2}{r_-} d r_- \Big) = 2 \pi^2 ( P_+ r_+^3 d r_+ + P_- r_-^3 d r_- ).
\ee
This gives the motivation for us to introduce the expressions of radial pressures as
$P_\pm \equiv T^r_{\ r} (r_\pm) = - \frac{3}{8 \pi G} \frac{r_\mp^2}{r_\pm^4}$. Just like the charged black holes in ref.~\cite{Kwon:2013dua}, $P_\pm$ is also proportional to the square of the electromagnetic parameter.

On the other hand, in order to incorporate the variation of a relevant quantity with respect to the coupling $\a$,
we define a total variational operator as $\wt d \equiv d + d_\a$.
For example, we have
\bea
\wt d S_+ &\equiv& d S_+ + d_\a S_+ = \f{3\pi^2}{2G} ( r_+^2 + 4\a ) d r_+ + \f{6\pi^2}{G} r_+ d \a,  \\
\wt d E_+ &\equiv& d E_+ + d_\a E_+ = \f{3\pi}{4G} ( r_+ d r_+ + d \a ).
\eea
Now we see that eq.~(\ref{m2}) turns into the precise form of the first law of black hole thermodynamics,
\be   \lab{lbh}
- d M + T_+ \wt d S_+ + \Phi_+ d q + \Theta_+ d \a= 0.
\ee
Here the potential conjugate to the Gauss-Bonnet coupling is $\Theta_+ = \f{3\pi}{4 G} (1 - 8 \pi r_+ T_+)$
and $T_+ d S_+$ can be rewritten as $\f{\k}{8\pi G} d A_+$.
For the sake of aesthetics, one can of course deduce an equivalent formula:
$ - \wt d M + T_+ \wt d S_+ + \Phi_+ \wt d q + \Theta_+ \wt d \a= 0 $.
It is worth mentioning that we can also obtain the first law at the inner horizon by applying the same method. 

We note that the extended first law (including variation with respect to the Gauss-Bonnet coupling) for the Lovelock gravity has been derived from first principles in ref.~\cite{Kastor:2010gq} and the result, together with the Smarr formula, has been utilized~\cite{Castro:2013pqa} to investigate some issues related to the universality of the product of horizon areas, where
a set of relations between thermodynamic potentials $\Theta_\pm$ has also been obtained.

Here it is necessary to make a comparison of our treatment with the earlier computation. 
We have adopted the method of ref.~\cite{Kwon:2013dua} to obtain the same extended first law as that of ref.~\cite{Castro:2013pqa} for Lovelock black holes. Our procedure is different from that of ref.~\cite{Kastor:2010gq}. Moreover, the first law at the inner horizon can be derived analogously. Therefore, it is  interesting that the same goal is reached by two different means.

%

\section{$f(R)$ black holes: constant curvature case}  \lab{sec3}

In this section, we turn to the investigation of charged black holes in the $f(R)$ gravity.
From the action
\be
S = \f{1}{16\pi G} \int d^4 x \sr{-g} \Big( R + f(R) \Big) + S_m, \label{fraction}
\ee
a class of $f(R)$-Maxwell black holes imposed by a constant curvature scalar $R=R_0$ can be obtained~\cite{Moon:2011hq,Sheykhi:2012zz},
\be
ds^2 = - N(r) dt^2 + \f{1}{N(r)} dr^2 + r^2 ( d\th^2 + \sin\th^2 d\phi^2 ),
\ee
\be     \lab{fnr}
N(r) = 1 - \f{2\m}{r} + \f{q^2}{r^2} \f{1}{1+f'(R_0)} - \f{R_0}{12} r^2 = - \f{R_0}{12 r^2} \prod_{i=1}^{4} (r-r_i)
, \qquad F_{tr} = \f{q}{r^2},
\ee
where $\m$ is a parameter related to the black hole mass.
From the action (eq.~(\ref{fraction})), 
we can derive the relevant component of field equations as follows:
\be   \lab{frr}
\f{1 + f'}{r_i} \Big( \f{2\m}{r_i^2} - \f{2q^2}{1+f'} \f{1}{r_i^3} - \f{R_0}{6} r_i - \f{1}{r_i} \Big)
- \f{1}{2} ( f - R_0 f' ) = 8\pi G P_i.
\ee
Since $N(r_i) = 0$, we have
\be                    \lab{dm}
\m = \f{r_i}{2} \Big( 1 + \f{q^2}{r_i^2} \f{1}{1+f'(R_0)} - \f{R_0}{12} r_i^2 \Big).
\ee
Noting that here the volume is $V_i=\f{4\pi}{3}r_i^3$ and multiplying by $\f{r_i^2}{2G}dr_i$ on both sides of eq.~(\ref{frr}), we arrive at the following equation,
\be     \lab{fr2}
\f{1+f'}{2G} \Big( 1- \f{q^2}{r_i^2} \f{1}{1+f'} - \f{R_0}{4} r_i^2 \Big) dr_i- \f{1+f'}{2G} dr_i - \f{1}{4G} ( f - R_0 f') r_i^2 dr_i = P_i d V_i.
\ee
Thus, it has the form of the first law of thermodynamics,
\be    \lab{ffl}
T_i d S_i - d E_i - T_i d \bar S_i = P_i d V_i,
\ee
where the black hole parameters are
\be
T_i = \f{1}{4\pi r_i} \Big( 1- \f{q^2}{r_i^2} \f{1}{1+f'} - \f{R_0}{4} r_i^2 \Big), \qquad S_i 
= \f{\pi r_i^2}{G} (1+f'),  \qquad E_i = \f{r_i}{2G} (1+f'), \label{tse}
\ee
where $E_i$ is just the Misner-Sharp energy.


As noted in ref.~\cite{Akbar:2006mq}, see also ref.~\cite{Eling:2006gr}, the additional entropy term\footnote{It is easy to get
$d \bar S_i = \frac{\pi (f-R_0 f')}{G} \frac{r^3_i}{ 1 - \frac{q^2}{r^2_i (1+f')} - \frac{R_0}{4} r^2_i} dr_i$ when we compare eq.~(\ref{fr2}) with eq.~(\ref{ffl}) and consider the expression of $T_i$ in eq.~(\ref{tse}). This formula  
coincides with that given by ref.~\cite{Akbar:2006mq}.} $d \bar S_i$ is the entropy production term in non-equilibrium thermodynamics. That is, when the $f(R)$ higher derivative term is included in the action of $f(R)$ gravity, the horizon thermodynamics will become non-equilibrium.
In this case, the entropy balance law needs to be modified~\cite{Eling:2006gr,Chirco2010} to derive the $f(R)$ gravity field equations from the thermodynamical prescription. 
The extra irreversible entropy production term can be interpreted as the bulk viscosity, and has its origin in the nonzero expansion of the null geodesics comprising the horizon. However, through a more general definition of local entropy, the reversible spacetime thermodynamics can still be applied~\cite{Elizalde2008} to this non-equilibrium case.
The original first law cannot be sustained unless the extra entropy production term $d \bar S_i$ is taken into account to balance the inequality.
In the limit of Einstein's gravity, this extra term will disappear.
On the other hand, if the energy term is redefined as $d \bar E_i=d E_i+\f{1}{4G} ( f - R_0 f') r_i^2 dr_i$, 
the first law eq.~(\ref{ffl}) turns into the following form,
\be
T_i d S_i - d \bar E_i = P_i d V_i, \label{frpdv}
\ee
which is exactly same as that in the equilibrium circumstance.

From the expression of $N(r)$ in eq.~(\ref{fnr}), we obtain some useful relations as follows:
\bea
\sum_{i=1}^{4} r_i = 0, \quad \prod_{i=1}^{4} r_i &=& - \f{12}{R_0} \f{q^2}{1+f'},  \lab{fre1}  \\
r_1 r_2 + r_2 r_3 + r_3 r_4 + r_4 r_1 + r_1 r_3 + r_2 r_4 &=& - \f{12}{R_0}, \lab{fre2} \\
r_1 r_2 r_3 + r_2 r_3 r_4 + r_3 r_4 r_1 + r_1 r_2 r_4 &=& \f{24\m}{R_0}, \lab{fre3}  \\
\sum_{i=1}^{4} r_i^2 d r_i &=& -\f{24}{R_0} d \m.  \lab{fre4}
\eea
Eq.~(\ref{fre4}) can easily be obtained from the helpful formula provided in ref.~\cite{Du:2014gr}, where it has been proved that the roots $r_i$  $(i=1, 2, \cdots, m)$ of the following polynomial
\be
a_m r^m + a_{m-1} r^{m-1} + .... + a_0 r^0 = 0,
\ee
satisfy a simple formula,
\be \lab{sn}
s_n=- \f{1}{a_m} \sum_{i=0}^{m-1} s_{n-m+i} a_i,
\ee
where $s_n \equiv \sum_{i=1}^m r_i^n$.
Note that $s_{n-m+i}=0$ for $n-m+i<0$, and $s_{n-m+i}=n$ for $n-m+i=0$.
For our case, the positions of horizons are determined by 
the equation $N(r)=0$ which can be put into a standard form as
\be
\f{R_0}{12} r^4 - r^2 + 2\m r - \f{q^2}{1+f'} = 0.
\ee
So we have $a_4=\f{R_0}{12}$, $a_3=0$, $a_2=-1$, $a_1=2\m$, and $a_0=-\f{q^2}{1+f'}$.
Substituting these coefficients into eq.~(\ref{sn}) and setting $n=3$ and $m=4$, we get
\be
\sum_{i=1}^{4} r_i ^3 = -\f{72}{R_0} \m.
\ee
By taking derivative on both sides of the above equation, we recover eq.~(\ref{fre4}).

With this digression finished, we are ready to use eq.~(\ref{fr2}) and sum the four equations for $i=1, 2, 3, 4$ to obtain a central equation,
\bea
& &T_1 d S_1 - \f{1+f'}{2G} d r_1
- \f{q^2}{2G} \Big( \f{1}{r_2^2} d r_2 +  \f{1}{r_3^2} d r_3 +  \f{1}{r_4^2} d r_4 \Big) \nn \\
& &+ \f{1+f'}{2G} \Big(-\f{R_0}{4}\Big) \big(r_2^2 d r_2 + r_3^2 d r_3 + r_4^2 d r_4 \big) - \f{1}{4 G} ( f - R_0 f' ) \sum_{i=1}^{4} r_i^2 d r_i \nn \\
& &= \sum_{i=1}^{4} P_i d V_i,
\eea
where the radial pressure is $P_i \equiv T^r_{\ r} (r_i) = - \frac{q^2}{8\pi G r_i^4}$. Note that the energy-momentum tensor of $f(R)$ black holes takes the same form as that of Lovelock black holes, see its formulation under eq.~(\ref{lrr}), because both kinds of black holes have the same Maxwell charge. The difference between them lies in metrics, which gives rise to different nonzero components of electromagnetic tensors $F_{tr}$, see eqs.~(\ref{ncet}) and (\ref{fnr}).
After regarding $r_1$ as the outermost  position of event horizons and making some manipulation, we have
\be
T_1 d S_1 - \Big(\f{1+f'}{2G} - \f{1+f'}{2G} \f{R_0}{4} r_1^2-\f{q^2}{2G} \f{1}{r_1^2}\Big) d r_1 + \Big[ \f{3}{G} (1+f') + \f{6}{G R_0} (f-R_0f')\Big] d \m = 0.
\ee

On the other hand,  when $i=1$ eq.~(\ref{dm}) leads to
\be
d \m = \f{1}{2} \Big( 1 - \f{q^2}{r_1^2} \f{1}{1+f'} - \f{R_0}{4} r_1^2 \Big) d r_1 + \f{q}{r_1} \f{1}{1+f'} d q.
\ee
Combining the above two equations, we obtain
\be
T_1 d S_1 + \f{q}{G r_1} d q + \Big[\f{2}{G}(1+f')  + \f{6}{G R_0} (f- R_0 f')\Big] d \m = 0.
\ee
By recalling the black hole parameters, i.e., the electric potential $\Phi_i$ on the $r_i$ horizon and the electric charge $Q$,
\be
\Phi_i = \sr{\f{2}{G}} \f{q}{r_i} \sr{1+f'}, \qquad Q = \f{1}{\sr{2G}} \f{q}{\sr{1+f'}},
\ee
and modifying the mass parameter,
\be
M = \f{\m}{G} (1+f') \longrightarrow \wt M = - \f{2\m}{G} \Big[ (1+f') + \f{3}{R_0} (f- R_0 f')\Big],
\ee
we finally derive the first law of black hole thermodynamics,
\be  \lab{dM}
- d \wt M + T_1 d S_1 + \Phi_1 d Q = 0.
\ee
Only when $f' - \f{2}{R_0} f - 1 = 0$ can we have $\wt M = M$.
This requires that $f(R) = C e^{\f{2R}{R_0}} - \f{R_0}{2}$, where $C$ is a constant.

We make some comments.
First, the physical interpretation of the new mass parameter $\wt M$ is unclear at present.
Without the introduction of $\wt M$ the expected first law could not be reproduced even if the non-equilibrium part were discarded.
Second, it is puzzling\footnote{In the case of two horizons (see eq.~(\ref{lre})) of the 5-dimensional Lovelock black holes, the both horizons have been utilized to obtain the first law.} that only two of the five relations given in eqs.~(\ref{fre1})-(\ref{fre4}) have been used in the derivation of the first law.
The phenomenon brings one to look for an alternative derivation.
Third,
it is unclear how to associate a proper interpretation with the thermodynamic first law at virtual horizons
due to the restrictions of the method itself.
Finally,
one can make use of the analogous method to study other black holes in the $f(R)$ gravity as done in
refs.~\cite{delaCruzDombriz:2009et,Sebastiani:2010kv,Hendi:2011eg}.

\section{Conclusion}

In ref.~\cite{Kwon:2013dua} an interesting property is discovered that
the contributions of inner horizons should be considered when one derives the first law of black hole thermodynamics.
The essential step is to sum the equations corresponding to the first law of thermodynamics at all horizons.
By applying this method,
we have studied a 5-dimensional charged Lovelock black hole and a 4-dimensional $f(R)$-Maxwell black hole imposed by a constant curvature scalar.
More general black holes in the Lovelock gravity and $f(R)$ gravity can be analyzed similarly.

This work may be extended along the following ways.
Firstly,
one may attempt to rigorously prove the property found by ref.~\cite{Kwon:2013dua}.
In this aspect, 
the investigation in ref.~\cite{Kothawala:2009kc} may be helpful where the near horizon symmetries of the
Einstein tensor are used to demonstrate the thermodynamic interpretation of the field equations near the horizon.
Secondly,
it is interesting to generalize this method to the case of nonzero variation of the cosmological constant.
We note that since the work of ref.~\cite{Kastor:2009wy},
a lot of efforts have been devoted to study this kind of extended first law, see, for instance, ref.~\cite{Cvetic:2011}, and its relevant phase transitions~\cite{Altamirano:2014tva}.
Thirdly, as
a previous work pointed out~\cite{Son:2013eea}, the pressure plays a complementary role in the black hole thermodynamics.
Curiously, the literature focusing on the thermodynamic volume~\cite{Altamirano:2014tva} involves a term like $V d P$ rather than $P d V$.
We note that $P d V$ appears in eqs.~(\ref{sum}) and (\ref{frpdv}), but does not appear in the final expressions of black hole thermodynamics  eqs.~(\ref{lbh}) and (\ref{dM}), and that  $V d P$ actually plays no role in the black hole thermodynamics investigated in the present paper.
Both terms may have some connection to the method proposed in ref.~\cite{Kwon:2013dua}.

\section*{Acknowledgments}

The authors would like to thank the anonymous referee for the helpful comments that indeed improve this work greatly. This work was supported in part by the National Natural Science Foundation of China under grant No.11175090 and by the Ministry of Education of China under grant No.20120031110027.


\end{document}